\documentclass[10pt,twocolumn,twoside]{IEEEtran} 

\usepackage[cmex10]{amsmath}
\usepackage{amssymb}
\usepackage{fixltx2e}
\usepackage[T1]{fontenc}
\usepackage{url}
\usepackage{dsfont}
\usepackage{mathtools}
\usepackage{color}
\usepackage{graphicx}
\usepackage{bm}

\usepackage{tabularx}

\newtheorem{thm}{Theorem}
\newtheorem{prop}[thm]{Proposition}  			   
\newtheorem{cor}[thm]{Corollary}     		   	   
\newtheorem{rem}[thm]{Remark}

\newcommand{\R}{\mathds{R}}

\newcommand{\N}{\mathds{N}}
\newcommand{\Q}{\mathds{Q}}

\newcommand{\cS}{\mathcal{S}}
\newcommand{\cC}{\mathcal{C}}
\newcommand{\st}{\quad\text{s.t.}\quad}
\newcommand{\norm}[1]{\lVert {#1} \rVert}
\newcommand{\Norm}[1]{\left\lVert {#1} \right\rVert}
\newcommand{\abs}[1]{\lvert {#1} \rvert}

\newcommand{\define}{\coloneqq}

\newcommand{\NP}{\textsf{NP}}
\newcommand{\DTIME}{\textsf{DTIME}}
\renewcommand{\P}{\textsf{P}}

\DeclareMathOperator{\rank}{rank}

\newcommand{\M}{\bm{M}}
\newcommand{\A}{\bm{A}}
\newcommand{\matP}{\bm{P}}
\newcommand{\B}{\bm{B}}
\newcommand{\X}{\bm{X}}
\newcommand{\Y}{\bm{Y}}
\newcommand{\D}{\bm{D}}
\newcommand{\x}{\bm{x}}
\newcommand{\y}{\bm{y}}

\begin{document}
\title{
  On the Computational Intractability of\\
  Exact and Approximate Dictionary Learning}%
%
\author{Andreas M. Tillmann~\IEEEmembership{}
\thanks{A. M. Tillmann is with the Research Group Optimization at TU
  Darmstadt, Dolivostr. 15, 64293 Darmstadt, Germany (phone:
  +49-6151-1670868, e-mail: tillmann@mathematik.tu-darmstadt.de).}%
\thanks{This work has been accepted by the IEEE for publication. Copyright may be transferred without notice, after
    which this version may no longer be accessible.}%
}%
\markboth
{On the Computational Intractability of Exact and Approximate Dictionary Learning}
{On the Computational Intractability of Exact and Approximate Dictionary Learning}


\maketitle

\begin{abstract}
  The efficient sparse coding and reconstruction of signal vectors via
  linear observations has received a tremendous amount of attention
  over the last decade. In this context, the automated learning of a
  suitable basis or overcomplete dictionary from training data sets of
  certain signal classes for use in sparse representations has turned
  out to be of particular importance regarding practical signal
  processing applications. Most popular dictionary learning algorithms
  involve \NP-hard sparse recovery problems in each iteration, which
  may give some indication about the complexity of dictionary learning
  but does not constitute an actual proof of computational
  intractability. In this technical note, we show that learning a
  dictionary with which a given set of training signals can be
  represented as sparsely as possible is indeed
  \NP-hard. 
  Moreover, we also establish hardness of approximating the solution
  to within large factors of the optimal sparsity level. Furthermore,
  we give \NP-hardness and non-approximability results for a recent
  dictionary learning variation called the sensor permutation
  problem. Along the way, we also obtain a new non-approximability
  result for the classical sparse recovery problem from compressed
  sensing.
\end{abstract}



%
\IEEEpeerreviewmaketitle

\section{Introduction}
\IEEEPARstart{A}{s} a central problem in compressed sensing (CS)
\cite{D06,FR13,EK12}, the task of finding a sparsest exact or
approximate solution to an underdetermined system of linear equations
has been a strong focus of research during the past decade. Denoting
by~$\norm{\x}_0$ the so-called $\ell_0$-norm, i.e., the number of
nonzero entries in~$\x$, the sparse recovery problem reads
\begin{equation}\label{prob:P0}
  \min\,\norm{\x}_0\quad\st\quad \norm{\D\x-\y}_2\leq\delta,\tag{$\text{P}_0^\delta$}
\end{equation}
for a given matrix $\D \in \R^{m \times n}$ with $m \leq n$ and an
estimate $\delta\geq 0$ of the amount of error contained in the
measurements $\y\in\R^m$. Both the noisefree problem
$(\text{P}_0)\define(\text{P}_0^0)$ and the error-tolerant variant
\eqref{prob:P0} with $\delta>0$ are well-known to be \NP-hard in the
strong sense, cf.~\cite[problem MP5]{GJ79} and \cite{N95}, and also
difficult to approximate~\cite{AK98,A99}.


Groundbreaking results from CS theory include qualificatory conditions
(on the dictionary $\D$ and the solution sparsity level) 
which yield efficient solvability of the generally hard
problems~\eqref{prob:P0} 
by greedy methods---e.g., orthogonal matching
pursuit (OMP) \cite{PRK93}---or (convex) relaxations such as 
Basis Pursuit 
\cite{CDS98} (
replacing the $\ell_0$-norm 
by the $\ell_1$-norm); see
\cite{FR13,EK12,T13} for overviews. Subsequently, numerous
optimization algorithms have been tailored to sparse recovery tasks,
and various types of dictionaries 
were shown or
designed to exhibit favorable recoverability properties. In
particular, the essential assumption of (exact or approximate) sparse
representability of certain signal classes using specific dictionaries
has been empirically verified in many practical signal processing
applications; for instance, natural images are known to admit sparse
approximations over discrete cosine or wavelet bases \cite{FR13}.

Nevertheless, a predetermined setup typically cannot fully capture the
true structure of real-world signals; thus, using a \emph{fixed}
dictionary $\D$ naturally restricts the achievable sparsity levels of
the representations. Indeed, the simultaneous search for both
dictionary and sparse representations of a set of training
signals---commonly referred to as \emph{dictionary learning}
---was demonstrated to allow for
significantly improved sparsity levels using the learned dictionary
instead of an analytical, structured or random one. Successful
applications of dictionary learning include diverse tasks such as
image inpainting \cite{AEB06b,MBPS10}, denoising \cite{EA06,BSF13} and
deblurring \cite{CDMBP11}, or audio and speech signal representation
\cite{YBD08,JP11}, to name but a few.

Somewhat informally, the dictionary learning (DL) problem can be stated as:
  {\itshape Given a collection of training
  data vectors $\y^1,\dots,\y^p\in\R^m$ and a positive integer~$n$, 
  find a matrix
  $\D\in\R^{m\times n}$ that allows for the sparsest possible
  representations~$\x^j$ such that $\D\x^j= \y^j$ (for all $j$). }
This task can be formalized in different ways, and there exist many
variants seeking dictionaries with further properties such as
incoherence \cite{BP13} or union-of-bases \cite{LGBB05}; see also,
e.g., \cite{RZE10,P07,MBPS10}. Moreover, several DL
algorithms have been developed over the past years; the frequently
encountered hard sparse recovery subproblems are typically treated by
classical methods from CS. 
We refer to \cite{EAH99,A06,AEB06,GS10,KDMRELS03,MBPS10,R13,ABGM14} (and
references therein) for a broader overview of well-established
DL techniques and some more recent results.

In this paper, we are concerned with the computational complexity of
dictionary learning. Due to its combinatorial nature, it is widely
believed to be a very challenging
problem, 
but to the best of our knowledge, a formal proof of this
intractability claim was missing. We contribute to the theoretical
understanding of the problem by providing an \NP-hardness proof as
well as a strong non-approximability result for DL, see
Section~\ref{sect:DLcomplexity}. Furthermore, we prove \NP-hardness
and non-approximability of an interesting new DL variant---the
\emph{sensor permutation problem}, where the sought dictionary is
constrained to be related to a given sensing matrix via unknown row
permutations; see Section~\ref{sect:SPPcomplexity} for the details. As
a byproduct, 
we also obtain a new \NP-hardness of approximation result for the
sparse recovery problem~\eqref{prob:P0}.

\begin{rem}\label{rem:NP}
  Recall that \NP-hardness implies that no polynomial-time solution
  algorithm can exist, under the most-widely believed theoretical
  complexity assumption that \P$\neq$\NP~\cite{GJ79}. Further,
  \emph{strong} \NP-hardness can be understood, in a nutshell, as an
  indication that a problem's intractability does not depend on
  ill-conditioning of the input coefficients. This additionally
  implies that (unless \P$=$\NP) there cannot exist a
  pseudo-polynomial-time exact algorithm and not even a fully
  polynomial-time approximation scheme (FPTAS), i.e., an algorithm
  that solves a minimization problem within a factor of~$(1 +
  \varepsilon)$ of the optimal value in polynomial time with respect
  to the input size and $1/\varepsilon$, see
  \cite{GJ79}. 
For a thorough and detailed treatment of complexity theory, we refer
to~\cite{GJ79,KV11}.
\end{rem}

\section{The Complexity of Dictionary Learning}\label{sect:DLcomplexity}
As mentioned in the introduction, different philosophies or goals lead
to different formulations of dictionary learning problems, which are
usually captured by the general form
\begin{equation}\label{prob:DLgen}
  \min_{\D,\X} f(\D,\X;\Y) + g(\D) + h(\X),
\end{equation}
where the variables are the dictionary $\D\in\R^{m\times n}$ (for an a
priori chosen $n$) and the matrix $\X\in\R^{n\times p}$, whose columns 
are the representation vectors $\x^j$ of the given training
signals $\y^j$ (w.r.t. the linear model assumption
$\D\x^j\approx\y^j$), collected in $\Y\in\R^{m\times p}$ as its
columns; 
the functions $f$, $g$ and $h$ express a data fidelity
term, and constraints or penalties/regularizers for the dictionary and
the representation coefficient vectors, resp.

In the (ideal) noiseless case, the usual approach (see, e.g.,
\cite{AEB06b,P07,GS10}) is
\begin{equation}\label{prob:myDL}
  \min_{\D,\X} \Norm{\X}_0 \st \D\X=\Y,
\end{equation}
which fits the framework \eqref{prob:DLgen} by setting
$f(\D,\X;\Y)\define \chi_{\{\D\X=\Y\}}(\D,\X)$ (where $\chi$ is the
indicator function, i.e., $f(\D,\X;\Y)=0$ if $\D\X=\Y$ and $\infty$
otherwise), $g(\D)\define 0$ and $h(\X)\define\Norm{\X}_0$ (
extending the usual notation to matrices, $\Norm{\X}_0$ counts the
nonzero entries in $\X$). This problem is~a natural extension of
$(\text{P}_0)$, and can also be seen as a matrix-factorization
problem. 
To mitigate 
scaling ambiguities, one often sets $g(\D)\define
\chi_{\{\norm{\D_j}_2\leq 1~\forall\,j=1,\dots,n\}}(\D)$, i.e., the
columns of~$\D$ are required to have bounded norms;
cf.~\cite{KDMRELS03,YBD08}.

Note that if $n$ is not fixed a priori to a value smaller than $p$,
the dictionary learning task becomes trivial: Then, we could just take
$\D=[\y^1,\dots,\y^p]$ and exactly represent every $\y^i$ using only
one column. (Clearly, this also holds for variants which allow
representation errors, e.g., $\norm{\D\x^j-\y^j}_2\leq\delta$ for
some~$\delta>0$, or minimize such errors under hard sparsity limits
$\norm{\x^j}_0\leq k$ for some $k\geq 1$.) Thus, requiring $n<p$ is
hardly restrictive, in particular since the training data set
(and hence, $p$) is usually very large---intuitively, the more samples
of a certain signal class are available for learning the dictionary,
the better the outcome will be adapted to that signal class---and
with respect to storage aspects and efficient (algorithmic) applicability
of the learned dictionary, settling for a smaller number of dictionary
atoms is well-justified. Similarly, $m\leq n$ is a natural assumption,
since sparsity of the coefficient vectors is achieved via appropriate
representation bases or redundancy (overcompleteness) of the
dictionary; also, at least for large $p$, one can expect
$\rank(\Y)=m$, in which case $\rank(\D)=m\leq n$ becomes necessary to
maintain $\D\X=\Y$.

\subsection{\NP-Hardness}
As the following results show, finding a dictionary with which the
training signals can be represented with optimal sparsity is indeed a
computationally intractable problem.

\begin{thm}\label{thm:DLhardness}
  Solving the dictionary learning problem \eqref{prob:myDL} is
  \NP-hard in the strong sense, even when restricting $n=m$. 
\end{thm}
\begin{IEEEproof}
  We reduce from the \emph{matrix sparsification} (MS) problem: Given
  a full-rank
  matrix $\M\in\Q^{m\times p}$ ($m<p$), find a regular
  matrix $\B\in\R^{m\times m}$ such that $\B\M$ has as few nonzero
  entries as possible. (The full-rank assumption is not mandatory, but
  can be made w.l.o.g.: 
  If $\rank(\M)=k<m$, $m-k$ rows can be zeroed in polynomial time by
  elementary row operations, reducing the problem to sparsifying the
  remaining $k$-row submatrix.) 
  The MS problem was shown to be
  \NP-hard in \cite[Theorem~3.2.1]{McC83} (see also \cite{CP84,T13}),
  by a reduction from \emph{simple max cut}, cf. \cite[problem
  ND16]{GJ79}; since this reduction constructs a binary
  matrix (of dimensions polynomially bounded by the cut problem's
  input graph size), \NP-hardness of MS in fact holds in the
  strong sense, and we may even assume w.l.o.g. that
  $\M\in\{0,1\}^{m\times p}$.

  From an instance of MS, we obtain an equivalent instance of 
  \eqref{prob:myDL} as follows: Set $n\define m$ and let
  $\Y\define\M$. Then, the task~\eqref{prob:myDL} is to find
  $\D\in\R^{m\times m}$ and $\X\in\R^{m\times p}$ such that $\D\X=\Y$
  and $\Norm{\X}_0$ is minimal. (Note that the sought dictionary in
  fact constitutes a basis for $\R^m$, since $\M$ has full
  row-rank~$m$, thus requiring this of the dictionary as well, as
  discussed above.) Clearly, an optimal solution ($\D_*$, $\X_*$) of
  this dictionary learning instance gives an optimal solution
  $\B_*=\D_*^{-1}$ of MS, with $\B_*\M=\X_*$. It remains to note that
  the reduction is indeed polynomial, since the matrix inversion can
  be performed in strongly polynomial time by Gaussian elimination,
  cf. \cite{GLS93}. Thus, \eqref{prob:myDL} is strongly \NP-hard.
\end{IEEEproof}

\begin{rem}\label{rem:normConstraints}
  The above \NP-hardness result easily extends to variants of
  \eqref{prob:myDL} with the additional constraint that, for some
  constant~$c>0$, $\norm{\D_j}_2\leq c$ for all $j$, or
  $\Norm{\D}_{\text{F}}^2=\text{tr}(\D^\top\D)\leq c$ (as treated
  in~\cite{YBD08}): Since the discrete objectives are invariant to
  scaling in both the dictionary learning and the MS problem, there is
  always also an optimal $\D_*'$ (achieving the same number of
  nonzeros in the corresponding $\X_*'$) that obeys the norm
  constraints and yields an associated optimal solution
  $\B_*'=(\D_*')^{-1}$ of the MS problem. (Clearly, this argument
  remains valid for a host of similar norm constraints as well.)
\end{rem}

It is not known whether the decision version of the MS problem is
contained in \NP\ (and thus not only \NP-hard but \NP-complete)
\cite{McC83}. Similarly, we do not know if the decision problem
associated with 
\eqref{prob:myDL}---``given
$\Y\in\Q^{m\times p}$ and positive integers $k$ and~$n$, decide
whether there exist $\D\in\R^{m\times n}$ and $\X\in\R^{n\times p}$
such that $\D\X=\Y$ and $\Norm{\X}_0\leq k$''---is contained in \NP,
even in the square case $n=m$.

\subsection{Non-Approximability}
Since for \NP-hard problems, the existence of efficient
(polynomial-time) general exact solution algorithms is deemed
impossible, it is natural to search for good approximation methods. 
Indeed, virtually all well-known dictionary learning algorithms 
can be interpreted (in a vague sense) as ``approximation schemes''
since, e.g., the $\ell_0$-norm is convexified to the $\ell_1$-norm,
constraints may be turned to penalty terms in the objective
(regularization), etc. However, even disregarding the computational
costs of the algorithms, little is known about the quality of the
obtained approximations; several recent works along these lines
started investigating theoretical recovery properties and error
guarantees of dictionary learning algorithms, see, e.g.,
\cite{SWW12,AAJNT13,ABGM14}; in particular, \cite{AAJNT13} shows the
importantance of a good dictionary initialization.

The non-existence of an FPTAS (cf.
Remark~\ref{rem:NP}) 
itself does not generally rule out the existence of an efficient
algorithm with some constant approximation guarantee. However, we show
below that it is \emph{almost-\NP-hard} to approximate the dictionary
learning problem~\eqref{prob:myDL} to within large factors of the
optimal achievable sparsity of representations. Almost-\NP-hardness
means that no polynomial-time algorithm (here, to achieve the desired
approximation ratio) can exist so long as
\NP$\not\subseteq$\DTIME$(N^{\mathrm{poly}(\log N)})$, where~$N$ 
measures 
the input size (usually, dimension);
cf.~\cite{ABSS97}. 
This complexity assumption is stronger than \P$\neq$\NP, but also
firmly believed (cf., e.g.,
\cite{Trev13,BH92,IP01}); it essentially amounts to the claim that not
all \NP-hard problems admit a quasi-polynomial-time
deterministic 
solution algorithm. Many of the best known non-approximability
results are based on this assumption (see, e.g., \cite{AL96,Trev13}).

\begin{thm}\label{thm:DLnon-approx}
  For any $\varepsilon>0$, the dictionary learning
  problem~\eqref{prob:myDL} cannot be approximated within a factor of
  $2^{\log^{1-\varepsilon} m}$ in polynomial time, unless
  \NP$\subseteq$\DTIME$(m^{\mathrm{poly}(\log m)})$.
\end{thm}
\begin{IEEEproof}
  In~\cite{GN10}, almost-\NP-hardness of approximating the optimal
  value of the matrix sparsification problem\footnote{The MS problem
    in \cite{GN10} is defined precisely in transposed form compared to
    the present paper, i.e., there, one seeks to sparsify a full-rank
    matrix with more rows than columns by right-multiplication with an
    invertible matrix.} to within a factor of $2^{\log^{1/2-o(1)} m}$
  (i.e., $2^{\log^{1/2-\varepsilon} m}$ for any~$\varepsilon>0$) was
  shown, based on results from \cite{ABSS97} for the problem of
  minimizing the number of violated equations in an infeasible
  linear equation system (see also \cite{AK98}, where this is called
  MinULR). A closer inspection of \cite[Section~3]{GN10} and
  \cite[Theorems~7 and~8]{ABSS97} reveals that this
  non-approximability result in fact holds up to factors of
  $2^{\log^{1-\varepsilon} m}$ for any $\varepsilon>0$. Since our
  reduction in the proof of Theorem~\ref{thm:DLhardness} is cost-preserving, 
  this result carries over directly.
\end{IEEEproof}

This shows that it is extremely unlikely to efficiently learn a
dictionary that yields provably good approximations of the sought
sparse representations of the training data.
\begin{rem}\label{rem:ApproxRemark}
  The extensions of problem \eqref{prob:myDL} that incorporate norm
  bounds on $\D$ 
  are equally hard to approximate since the respective objectives do
  not differ from the original matrix sparsification problem's
  objective, 
  cf. Remark~\ref{rem:normConstraints}. 
  Moreover, the chain of reductions ending in the above result and
  starting with \cite[Theorem~7]{ABSS97}, maintains a polynomial
  relationship between the dimensions (here, $m$ and $p$);
  thus, almost-\NP-hardness also holds for approximation to within
  $2^{\log^{1-\varepsilon} p}$. 
\end{rem}
\begin{rem}
  One may also be interested in learning an \emph{analysis} dictionary
  $\bm{\Omega}$, minimizing $\norm{\bm{\Omega}\x}_0$ (for given $\x$),
  see, e.g., \cite{RPE13,YNGD11}. Imposing that $\bm{\Omega}$ has full
  rank excludes the trivial solution $\bm{\Omega}=0$ and, in fact, the
  square case then is completely equivalent to the MS problem, showing
  strong \NP-hardness and almost-\NP-hardness of approximation for
  analysis dictionary learning; Remarks~\ref{rem:normConstraints} and
  \ref{rem:ApproxRemark} apply similarly.
 \end{rem}

  \section{Sparse Coding with Unknown Sensor
    Locations}\label{sect:SPPcomplexity}
  Recently, an interesting new problem was introduced in \cite{EBDG14}
  and dubbed the ``sensor permutation problem'' (for short, SP). Here,
  it is assumed that the dictionary $\D$ is known up to a permutation
  of its rows, and one wishes to obtain the sparsest representations
  of the observations $\Y$ achievable via permuting these rows---or
  equivalently, the measurement entries. This approach can model,
  e.g., faulty wiring in the measurement system setup
  \cite{EBDG14}. Formally, the SP problem can be stated as
  \begin{equation}\label{prob:SPP}
    \min_{\matP,\X} \Norm{\X}_0 \st \A\X=\matP\Y,~\matP\in\mathcal{P}_m,
  \end{equation}
  where $\A\in\R^{m\times n}$ is a known dictionary, $\Y\in\R^{m\times
    p}$ and $\mathcal{P}_m\define\{\matP\in\{0,1\}^{m\times m}:\norm{\matP}_1=\norm{\matP}_\infty=1,\,\matP^\top\matP=I\}$ denotes the set of all $m\times m$
  permutation matrices. 
  \eqref{prob:SPP} can also be seen as a special case of the general
  dictionary learning framework~\eqref{prob:DLgen}, with
  $f(\D,\X;\Y)=\chi_{\{\D\X=\Y\}}(\D,\X)$,
  $g(\D)=\chi_{\{\D=\matP^\top\A\,:\,\matP\in\mathcal{P}_m\}}(\D)$ and
  $h(\X)=\Norm{\X}_0$.

  As our following results show, the sensor permutation problem is
  computationally intractable, even for ``nice'' input that does not
  contain numbers of highly varying sizes.

\begin{thm}\label{thm:SPPhardness}
  Problem \eqref{prob:SPP} is \NP-hard in the strong sense, even if
  $\A$ and $\Y$ are binary and $p=1$. Moreover, for any
  $\alpha\in(0,1)$ and any $\varepsilon>0$, there is no
  polynomial-time algorithm to approximate \eqref{prob:SPP} within a
  factor of $(1-\alpha)\ln(m)$ unless \P$=$\NP, or to within a
  factor of $2^{\log^{1-\varepsilon} m}$ unless
  \NP$\subseteq$\DTIME$(m^{\mathrm{poly}(\log m)})$. These results
  remain valid when $\A\X=\matP\Y$ is relaxed to
  $\Norm{\A\X-\matP\Y}_2\leq\delta$ for $0<\delta\in\R$, and/or $m$ is
  replaced by $n$.
\end{thm}

For the proof, recall the well-known strongly \NP-hard Set Cover
problem (SC, cf. \cite[problem SP5]{GJ79}): ``Given a set $\cS$ 
and a collection $\cC$ of 
subsets of $\cS$, find a
\emph{cover} of minimum cardinality, i.e., a subcollection~$\cC'$ of
as few sets from $\cC$ as possible such that
$\bigcup_{C\in\cC'}C=\cS$''. A cover~$\cC'$ is called \emph{exact} if
$C\cap D=\emptyset$ for all $C,D\in\cC'$ (in other words, if every
element of $\cS$ is contained in exactly one set from
$\cC'$).

We will employ the following very recent result: 
\begin{prop}[{\cite[Theorem~2]{M14}}]\label{prop:M14}
  For every $0<\alpha<1$, there exists a polynomial-time
  reduction from an arbitrary instance of the strongly \NP-complete
  satisfiability problem (SAT, cf. \cite[problem LO1]{GJ79}) to an
  SC instance ($\cS$,\,$\cC$) with a parameter $k\in\N$ such that
  if the input SAT instance is satisfiable, there is an exact cover of
  size~$k$ (and no smaller covers), whereas otherwise, every cover has
  size at least $(1-\alpha)\ln(\abs{\cS})\,k$.
\end{prop}

Recall also that, for any $\varepsilon>0$, approximating the sparse
recovery problem \eqref{prob:P0} (with any $\delta\geq 0$) to within
factors $2^{\log^{1-\varepsilon} n}$ is almost-\NP-hard, by
\cite[Theorem~3]{A99}. (In fact, although it clearly goes through for
$\delta=0$ as well, \cite{A99} states the proof of this only for
$\delta>0$, because the corresponding result for (P$_0$) had already
been shown in \cite{AK98} before.) The proof of \cite[Theorem~3]{A99}
is based on a special SC instance construction from \cite{BGLR93} (see
also \cite[Proposition~6]{ABSS97}) similar to that from
Proposition~\ref{prop:M14}.

\begin{rem}\label{rem:nvsm}
  In the special SC instances underlying the above results, it
  holds that $\abs{\cC}$ and $\abs{\cS}$ are polynomially related, so
  that all non-approximability results stated in this section also
  hold with $m$ ($=\abs{\cS}$) replaced by $n$ ($=\abs{\cC}$).
\end{rem}

We are now ready to prove the main result of this section.
\begin{IEEEproof}[Proof of Theorem~\ref{thm:SPPhardness}]
  Let $(\cS,\,\cC,\,k,\,\alpha)$ be a Set Cover instance as in
  Proposition~\ref{prop:M14}, and let $n=\abs{\cC}$, $m=\abs{\cS}$. 
  Following the proof of \cite[Theorem~3]{A99}, we first transform the
  task of finding a minimum-cardinality set cover to the sparse
  recovery problem~(P$_0$): Define $\D\in\{0,1\}^{m\times n}$ by
  setting $\D_{ij}=1$ if and only if the $i$-th element of $\cS$ is
  contained in the $j$-th set from $\cC$, and set
  $\y\define\mathds{1}$, i.e., the all-ones vector of length~$m$. It
  is easily seen that the support of every solution $\x$ of $\D\x=\y$
  induces a set cover (if some element was not covered, at least one
  row of the equality system would evaluate to $0=1$, contradicting
  $\D\x=\y$). Conversely, every \emph{exact} cover induces a solution
  of the same $\ell_0$-norm as the cover size (put $\x_C=1$ for the
  sets $C$ contained in the exact cover, and zero in the remaining
  components).
  Thus, if there is an \emph{exact} cover of size $k$, there is a
  $k$-sparse solution of $\D\x=\y$. Conversely, if all set covers have
  size at least $(1-\alpha)\ln(m)\,k$, then necessarily all
  $\x$ with $\D\x=\y$ have $\norm{\x}_0\geq
  (1-\alpha)\ln(m)\,k$ (because otherwise, the support of~$\x$
  would yield a set cover of size smaller than
  $(1-\alpha)\ln(m)\,k$). 


  This instance of (P$_0$) is now easily transformed into one of the
  sensor permutation problem \eqref{prob:SPP}: We set $\A\define\D$,
  $\Y\define\y$ (thus, $p=1$). 
  Now, since $\Y=\mathds{1}$, $\matP\Y=\Y$ for all
  $\matP\in\mathcal{P}_m$ and the choice of~$\matP$ has no influence
  on the solution. Thus, indeed, the SP problem~\eqref{prob:SPP} for
  these $\A$ and $\Y$ 
  has precisely the same solution value as the above-constructed
  instance of (P$_0$). Since solving the original Set Cover instance
  is (strongly) \NP-hard (by Proposition~\ref{prop:M14}), and all
  constructed numbers and their encoding lengths remain polynomially
  bounded by the input parameter $m$ (and $n$), this immediately shows
  the claimed strong \NP-hardness result. In fact, could we
  approximate, in polynomial time, the optimal solution value
  of~\eqref{prob:SPP} to within a factor of
  $(1-\alpha)\ln(m)$, then we could also decide the SAT
  instance underlying the SC problem from Proposition~\ref{prop:M14}
  in polynomial time, which is impossible unless \P$=$\NP. Therefore,
  for any $0<\alpha<1$, even approximating~\eqref{prob:SPP} to within
  factors $(1-\alpha)\ln(m)$ is \NP-hard. 

  For the second non-approximability result of
  Theorem~\ref{thm:SPPhardness}, it suffices to note that the
  construction above is cost-preserving and that the (P$_0$) instance
  in the proof of \cite[Theorem~3]{A99} also has $\y=\mathds{1}$.
  Hence, we can directly transfer the non-approximability properties,
  and conclude that there is no polynomial-time algorithm
  approximating \eqref{prob:SPP} to within factors
  $2^{\log^{1-\varepsilon} n}$ (for any $\varepsilon>0$), unless
  \NP$\subseteq$\DTIME$(n^{\mathrm{poly}(\log n)})$. 

  Finally, the above results extend to the noise-aware SP problem
  variant by treating the relaxed constraints
  $\Norm{\A\X-\matP\Y}_2\leq\delta$ for $\delta>0$ completely
  analogously to the proof of \cite[Theorem~3]{A99} (we omit the
  details) and, by Remark~\ref{rem:nvsm}, remain valid w.r.t. either
  $m$ or~$n$.
\end{IEEEproof}


\begin{rem}
  The decision version of~\eqref{prob:SPP} is easily seen to be in
  \NP\ (for rational input), and hence \NP-complete. 
\end{rem}

Note that the first part of the above proof yields a new
non-approximability result for sparse recovery:
\begin{cor}\label{cor:P0inapprox}
  For any $\alpha\in(0,1)$, it is \NP-hard to approximate
  \eqref{prob:P0} to within a factor of $(1-\alpha)\text{ln}(n)$.
\end{cor}
This complements the previously known results from
\cite[Theorem~7]{AK98} and \cite[Theorem~3]{A99}: For $n$ large enough
(and some fixed pair $\alpha$, $\varepsilon$),
$2^{\log^{1-\varepsilon} n}>(1-\alpha)\ln(n)$, but the
assumption \P$\neq$\NP\ is weaker than
\NP$\not\subseteq$\DTIME$(n^{\mathrm{poly}(\log n)})$.



\section{Concluding Remarks}
In this note, we gave formal proofs for \NP-hardness and
non-approximability of several dictionary learning problems. While
perhaps not very surprising, 
these results provide a complexity-theoretical justification for the
common approaches to tackle dictionary learning tasks by inexact
methods and heuristics without performance guarantees.

While preparing this manuscript, we became aware of a related result
presented at ICASSP 2014, see~\cite{RTL14}. In that work, the authors
claim \NP-hardness of approximating 
\begin{equation}\label{prob:otherDL}
  \min_{\D,\X} \Norm{\D\X-\Y}_{\text{F}}^2 \st \Norm{\X_j}_0\leq k~~\forall\,j=1,\dots,p,
\end{equation}
to within a given \emph{additive} error w.r.t. the objective (i.e.,
not within a \emph{factor} of the optimal value), for the case in which $\Y$
contains only two columns and $k$ is fixed to $1$.  Unfortunately,
\cite{RTL14} does not contain a proof, and at the time of writing, we
could not locate it elsewhere. 
Note also that, clearly, \eqref{prob:otherDL} is also a special case
of the general formulation~\eqref{prob:DLgen}---using
$f(\D,\X;\Y)=\Norm{\D\X-\Y}_{\text{F}}^2$,
$h(\X)=\chi_{\{\norm{\X_j}_0\leq k~\forall\,j\}}(\X)$ and
$g(\D)=0$---but that the results from the present paper and
from~\cite{RTL14} nevertheless pertain to different problems, both of
which are often referred to as ``dictionary learning''.

Future research closely related to the present work could include
investigating the potential use of matrix sparsification based
heuristics for dictionary learning purposes (e.g., when learning a
union-of-bases dictionary as in \cite{LGBB05}). 

Note also that the reduction from \cite{GN10} does not admit
transferring the \NP-hardness of approximating MinULR to within any
constant factor (see \cite[Theorem~5]{ABSS97}) to the MS
problem. (Similarly, the reduction to MS in 
\cite{McC83} apparently does not preserve approximation ratios.) Such
non-approximability results under the slightly weaker 
\P$\neq$\NP\ assumption hence remain open for
problem~\eqref{prob:myDL} (and its norm-constrained variants). Also,
the complexities of dictionary learning with $\ell_1$-objective and/or 
noise-awareness (e.g., constraints
$\norm{\D\X-\Y}_{\text{F}}\leq\delta$ for $\delta>0$) 
remain important open problems.

On the other hand, one may wish to focus on ``good news'', e.g., by
designing efficient approximation algorithms that give performance
guarantees not too much worse than our intractability thresholds, or
by identifying special cases which are notably easier to solve. Also,
it would be interesting to develop further ``hybrid algorithms'' that
combine relaxation methods and tools from combinatorial optimization,
such as the branch \& bound procedure from \cite{EBDG14}.



\section*{Acknowledgments}
The author would like to thank Yonina Eldar and Julien Mairal for
enticing him to look into dictionary learning, R\'{e}mi Gribonval for
bringing the sensor permutation problem to his attention, as well as
Imke Joormann, Marc Pfetsch and two anonymous referees for their
valuable comments on an earlier version of the manuscript.

\bibliographystyle{IEEEtran}

\begin{thebibliography}{10}
\providecommand{\url}[1]{#1}
\csname url@samestyle\endcsname
\providecommand{\newblock}{\relax}
\providecommand{\bibinfo}[2]{#2}
\providecommand{\BIBentrySTDinterwordspacing}{\spaceskip=0pt\relax}
\providecommand{\BIBentryALTinterwordstretchfactor}{4}
\providecommand{\BIBentryALTinterwordspacing}{\spaceskip=\fontdimen2\font plus
\BIBentryALTinterwordstretchfactor\fontdimen3\font minus
  \fontdimen4\font\relax}
\providecommand{\BIBforeignlanguage}[2]{{%
\expandafter\ifx\csname l@#1\endcsname\relax
\typeout{** WARNING: IEEEtran.bst: No hyphenation pattern has been}%
\typeout{** loaded for the language `#1'. Using the pattern for}%
\typeout{** the default language instead.}%
\else
\language=\csname l@#1\endcsname
\fi
#2}}
\providecommand{\BIBdecl}{\relax}
\BIBdecl

\bibitem{D06}
D.~L. Donoho, ``{Compressed Sensing},'' \emph{{IEEE Trans. Inform. Theory}},
  vol.~52, no.~4, pp. 1289--1306, 2006.

\bibitem{FR13}
S.~Foucart and H.~Rauhut, \emph{{A Mathematical Introduction to Compressive
  Sensing}}, ser. {Applied and Numerical Harmonic Analysis}.\hskip 1em plus
  0.5em minus 0.4em\relax Birkh\"{a}user, 2013.

\bibitem{EK12}
G.~Kutyniok and Y.~C. Eldar, Eds., \emph{{Compressed Sensing: Theory and
  Applications}}.\hskip 1em plus 0.5em minus 0.4em\relax Cambridge University
  Press, 2012.

\bibitem{GJ79}
M.~R. Garey and D.~S. Johnson, \emph{{Computers and Intractability. A Guide to
  the Theory of \NP-completeness}}.\hskip 1em plus 0.5em minus 0.4em\relax W.
  H. Freeman and Company, 1979.

\bibitem{N95}
B.~K. Natarajan, ``{Sparse Approximate Solutions to Linear Systems},''
  \emph{{SIAM J. Comput.}}, vol.~24, no.~2, pp. 227--234, 1995.

\bibitem{AK98}
E.~Amaldi and V.~Kann, ``{On the Approximability of Minimizing Nonzero
  Variables or Unsatisfied Relations in Linear Systems},'' \emph{{Theor.
  Comput. Sci.}}, vol. 209, no. 1--2, pp. 237--260, 1998.

\bibitem{A99}
E.~Amaldi, ``{On the complexity of designing compact perceptrons and some
  consequences},'' in \emph{{El. Proc. 5th Internat. Symp. on Artificial
  Intelligence and Math.}}, 1999.

\bibitem{PRK93}
Y.~C. Pati, R.~Rezaiifar, and P.~S. Krishnaprasad, ``{Orthogonal Matching
  Pursuit: Recursive Function Approximation with Applications to Wavelet
  Decomposition},'' in \emph{{Proc. 27th Ann. Asilomar Conference on Signals,
  Systems and Computers}}.\hskip 1em plus 0.5em minus 0.4em\relax IEEE Computer
  Society Press, 1993, vol.~1, pp. 40--44.

\bibitem{CDS98}
S.~S. Chen, D.~L. Donoho, and M.~A. Saunders, ``{Atomic Decomposition by Basis
  Pursuit},'' \emph{{SIAM J. Sci. Comput.}}, vol.~20, no.~1, pp. 33--61, 1998.

\bibitem{T13}
A.~M. Tillmann, ``{Computational Aspects of Compressed Sensing},'' Doctoral
  dissertation, TU Darmstadt, Germany, 2013.

\bibitem{AEB06b}
M.~Aharon, M.~Elad, and A.~M. Bruckstein, ``{On the uniqueness of overcomplete
  dictionaries, and a practical way to retrieve them},'' \emph{{Linear Algebra
  Appl.}}, vol. 416, no.~1, pp. 48--67, 2006.

\bibitem{MBPS10}
J.~Mairal, F.~Bach, J.~Ponce, and G.~Shapiro, ``{Online Learning for Matrix
  Factorization and Sparse Coding},'' \emph{{J. Mach. Learn. Res.}}, vol.~11,
  pp. 19--60, 2010.

\bibitem{EA06}
M.~Elad and M.~Aharon, ``{Image Denoising Via Sparse and Redundant
  Representations Over Learned Dictionaries},'' \emph{{IEEE Trans. Image
  Process.}}, vol.~15, no.~12, pp. 3736--3745, 2006.

\bibitem{BSF13}
S.~Beckouche, J.~L. Starck, and J.~Fadili, ``{Astronomical image denoising
  using dictionary learning},'' \emph{{Astron. Astrophys.}}, vol. 556, no.
  A132, 2013.

\bibitem{CDMBP11}
F.~Couzinie-Devy, J.~Mairal, F.~Bach, and J.~Ponce, ``{Dictionary Learning for
  Deblurring and Digital Zoom},'' arXiv:1110.0957 [cs.LG], 2011.

\bibitem{YBD08}
M.~Yaghoobi, T.~Blumensath, and M.~Davies, ``{Regularized Dictionary Learning
  for Sparse Approximation},'' in \emph{{Proc. EUSIPCO'08}}, 2008.

\bibitem{JP11}
M.~G. Jafari and M.~D. Plumbley, ``{Fast Dictionary Learning for Sparse
  Representation of Speech Signals},'' \emph{{IEEE J. Sel. Top. Signa.}},
  vol.~5, no.~5, pp. 1025--1031, 2011.

\bibitem{BP13}
D.~Barchiesi and M.~D. Plumbley, ``{Learning Incoherent Dictionaries for Sparse
  Approximation Using Iterative Projections and Rotations},'' \emph{{IEEE
  Trans. Signal Process.}}, vol.~61, no.~8, pp. 2055--2065, 2013.

\bibitem{LGBB05}
S.~Lesage, R.~Gribonval, F.~Bimbot, and L.~Benaroya, ``{Learning unions of
  orthonormal bases with thresholded singular value decompositon},'' in
  \emph{{Proc. IEEE ICASSP'05}}, 2005, vol.~5, pp. v/293--v/296.

\bibitem{RZE10}
R.~Rubinstein, M.~Zibulevsky, and M.~Elad, ``{Double Sparsity: Learning Sparse
  Dictionaries for Sparse Signal Approximation},'' \emph{{IEEE Trans. Signal
  Process.}}, vol.~58, no.~3, pp. 1553--1564, 2010.

\bibitem{P07}
M.~D. Plumbley, ``{Dictionary Learning for L1-Exact Sparse Coding},'' in
  \emph{{Independent Component Analysis and Signal Separation (Proc. ICA'07)}},
  ser. {Lect. Notes Comput. Sc.}\hskip 1em plus 0.5em minus 0.4em\relax
  Springer, 2007, vol. 4666, pp. 406--413.

\bibitem{EAH99}
K.~Engan, S.~O. Aase, and J.~H. Hus{\o}y, ``{Method of Optimal Directions for
  Frame Design},'' in \emph{{Proc. IEEE ICASSP'99}}, 1999, vol.~5, pp.
  2443--2446.

\bibitem{A06}
M.~Aharon, ``{Overcomplete Dictionaries for Sparse Representation of
  Signals},'' Ph.D. dissertation, {Technion -- Israel Institute of Technology},
  Haifa, Israel, 2006.

\bibitem{AEB06}
M.~Aharon, M.~Elad, and A.~M. Bruckstein, ``{K-SVD: An Algorithm for Designing
  of Overcomplete Dictionaries for Sparse Representations},'' \emph{{IEEE
  Trans. Signal Process.}}, vol.~54, no.~11, pp. 4311--4322, 2006.

\bibitem{GS10}
R.~Gribonval and K.~Schnass, ``{Dictionary Identification -- Sparse
  Matrix-Factorization via $\ell_1$-Minimization},'' \emph{{IEEE Trans. Inform.
  Theory}}, vol.~56, no.~7, pp. 3523--3539, 2010.

\bibitem{KDMRELS03}
K.~Kreutz-Delgado, J.~F. Murray, B.~D. Rao, K.~Engan, T.-W. Lee, and T.~J.
  Sejnowski, ``{Dictionary Learning Algorithms for Sparse Representation},''
  \emph{{Neural Comput.}}, vol.~15, no.~2, pp. 349--396, 2003.

\bibitem{R13}
A.~Rakotomamonjy, ``{Direct Optimization of the Dictionary Learning Problem},''
  \emph{{IEEE Trans. Signal Process.}}, vol.~61, no.~22, pp. 5495--5506, 2013.

\bibitem{ABGM14}
S.~Arora, A.~Bhaskara, R.~Ge, and T.Ma, ``{More Algorithms for Provable
  Dictionary Learning},'' arXiv:1401.0579 [cs.DS], 2014.

\bibitem{KV11}
B.~Korte and J.~Vygen, \emph{{Combinatorial Optimization. Theory and
  Algorithms}}, 5th~ed., ser. {Algorithms and Combinatorics}.\hskip 1em plus
  0.5em minus 0.4em\relax Springer, 2011, vol.~21.

\bibitem{McC83}
S.~T. McCormick, ``{A Combinatorial Approach to some Sparse Matrix Problems},''
  Ph.D. dissertation, Stanford University, CA, USA, 1983.

\bibitem{CP84}
T.~F. Coleman and A.~Pothen, ``{The Sparse Null Space Basis Problem},'' Cornell
  University, Ithaca, NY, USA, Tech. Rep. TR 84-598, 1984.

\bibitem{GLS93}
M.~Gr{\"o}tschel, L.~Lov{\'a}sz, and A.~Schrijver, \emph{{Geometric Algorithms
  and Combinatorial Optimization}}, 2nd~ed., ser. {Algorithms and
  Combinatorics}.\hskip 1em plus 0.5em minus 0.4em\relax Springer, 1993,
  vol.~2.

\bibitem{SWW12}
D.~A. Spielman, H.~Wang, and J.~Wright, ``{Exact Recovery of Sparsely-Used
  Dictionaries},'' \emph{{J. Mach. Learn. Res.}}, vol.~23, pp. 37.1--37.18,
  2012.

\bibitem{AAJNT13}
A.~Agarwal, A.~Anandkumar, P.~Jain, P.~Netrapalli, and R.~Tandon, ``{Learning
  Sparsely Used Overcomplete Dictionaries via Alternating Minimization},''
  arXiv:1310.7991 [cs.LG], 2013.

\bibitem{ABSS97}
S.~Arora, L.~Babai, J.~Stern, and Z.~Sweedyk, ``{The Hardness of Approximate
  Optima in Lattices, Codes, and Systems of Linear Equations},'' \emph{{J.
  Comput. Syst. Sci.}}, vol.~54, no.~2, pp. 317--331, 1997.

\bibitem{Trev13}
L.~Trevisan, ``{Inapproximability of Combinatorial Optimization Problems},'' in
  \emph{{Paradigms of Combinatorial Optimizations: Problems and New
  Approaches}}, V.~T. Paschos, Ed.\hskip 1em plus 0.5em minus 0.4em\relax John
  Wiley \& Sons, 2013, vol.~2, pp. 381--434.

\bibitem{BH92}
H.~Buhrman and S.~Homer, ``{Superpolynomial Circuits, Almost Sparse Oracles and
  the Exponential Hierarchy},'' in \emph{{Proc. FSTTCS 12}}, ser. Lect. Notes
  Comput. Sci.\hskip 1em plus 0.5em minus 0.4em\relax Springer, 1992, vol. 652,
  pp. 116--127.

\bibitem{IP01}
R.~Impagliazzo and R.~Paturi, ``{On the Complexity of $k$-SAT},'' \emph{J.
  Comput. Syst. Sci.}, vol.~62, pp. 367--375, 2001.

\bibitem{AL96}
S.~Arora and C.~Lund, ``{Hardness of Approximations},'' in \emph{{Approximation
  Algorithms for NP-hard Problems}}, D.~Hochbaum, Ed.\hskip 1em plus 0.5em
  minus 0.4em\relax PWS Publishing, 1996, pp. 399--446.

\bibitem{GN10}
L.-A. Gottlieb and T.~Neylon, ``{Matrix Sparsification and the Sparse Null
  Space Problem},'' in \emph{{Approximation, Randomization, and Combinatorial
  Optimization. Algorithms and Techniques (Proc. APPROX'10 and RANDOM'10)}},
  ser. {Lect. Notes Comput. Sci.}\hskip 1em plus 0.5em minus 0.4em\relax
  Springer, 2010, vol. 6302, pp. 205--218.

\bibitem{RPE13}
R.~Rubinstein, T.~Peleg, and M.~Elad, ``{Analysis K-SVD: A Dictionary-Learning
  Algorithm for the Analysis Sparse Model},'' \emph{{IEEE} Trans. Signal
  Process.}, vol.~61, no.~3, pp. 661--677, 2013.

\bibitem{YNGD11}
M.~Yaghoobi, S.~Nam, R.~Gribonval, and M.~E. Davies, ``{Analysis Operator
  Learning for Overcomplete Cosparse Representations},'' in \emph{{Proc.
  EUSIPCO'11}}, 2011.

\bibitem{EBDG14}
V.~Emiya, A.~Bonnefoy, L.~Daudet, and R.~Gribonval, ``{Compressed Sensing with
  Unknown Sensor Permutation},'' in \emph{{Proc. {IEEE} ICASSP'14}}, 2014, pp.
  1040--1044.

\bibitem{M14}
\BIBentryALTinterwordspacing
D.~Moshkovitz, ``{The Projection Games Conjecture and the NP-Hardness of
  ln\,$n$-Approximating Set-Cover},'' Preprint, 2014. [Online]. Available:
  \url{http://people.csail.mit.edu/dmoshkov/papers/set-cover/set-cover-full.pdf}
\BIBentrySTDinterwordspacing

\bibitem{BGLR93}
M.~Bellare, S.~Goldwasser, C.~Lung, and A.~Russell, ``{Efficient
  Probabilistically Checkable Proofs with Applications to Approximation
  Problems},'' in \emph{{Proc. 25th ACM Symp. Theory Comput.}}, 1993, pp.
  294--304.

\bibitem{RTL14}
M.~Razaviyayn, H.-W. Tseng, and Z.-Q. Luo, ``{Dictionary Learning for Sparse
  Representation: Complexity and Algorithms},'' in \emph{{Proc. {IEEE}
  ICASSP'14}}, 2014, pp. 5247--5251.

\end{thebibliography}


\end{document}